\documentclass{nature}
\usepackage{graphicx}
\usepackage{amsmath}
\usepackage{amssymb}
\usepackage{caption}

\usepackage{siunitx}
\usepackage[switch]{lineno}

\newcommand{\apj}{Astrophysical Journal}
\newcommand{\mnras}{Monthly Notices of Royal Astronomical Society}
\newcommand{\apjl}{Astrophysical Journal Letters}

\newcommand{\aap}{Astronomy \& Astrophysics}

\newcommand{\nat}{Nature}

\newcommand{\aplett}{Astrophysics Letters}
\newcommand{\araa}{Annual Review of Astron and Astrophys}

\title{Pulsar emission amplified and resolved by plasma lensing in an eclipsing binary}

\author{Robert Main$^{1,2,3}$, I-Sheng Yang$^{3,4}$, Victor Chan$^{1}$, Dongzi Li$^{3,5}$, Fang Xi Lin$^{3,5}$, Nikhil Mahajan$^{1}$, Ue-Li Pen$^{3,6,2,4}$, Keith Vanderlinde$^{1,2}$ \& Marten H. van Kerkwijk$^{1}$ }

\begin{document}

\maketitle

\vspace{12pt}

\blfootnote{
\begin{affiliations}
 \item Department of Astronomy and Astrophysics, University of Toronto, 50 St. George Street, Toronto, ON M5S 3H4, Canada
 \item Dunlap Institute for Astronomy and Astrophysics, University of Toronto, 50 St. George Street, Toronto, ON M5S 3H4, Canada
 \item Canadian Institute for Theoretical Astrophysics, University of Toronto, 60 St. George Street, Toronto, ON M5S 3H8, Canada
 \item Perimeter Institute for Theoretical Physics, 31 Caroline Street North, Waterloo, ON N2L 2Y5, Canada
 \item Department of Physics, University of Toronto, 60 St. George Street, Toronto, ON M5S 1A7, Canada
 \item Canadian Institute for Advanced Research, 180 Dundas St West, Toronto, ON M5G 1Z8, Canada
\end{affiliations}
}

\begin{abstract}

Radio pulsars scintillate because their emission travels through the
ionized interstellar medium via multiple paths, which interfere with
each other.  It has long been realized that the scattering screens
responsible for the scintillation could be used as ``interstellar
lenses'' to localize pulsar emission
regions\cite{lovelace70,backer75}.  Most scattering
screens, however, only marginally resolve emission
components, limiting results to statistical inferences and detections
of small positional shifts\cite{gwinn+12, johnson+12, pen+14}.
Since screens situated close to the source have better resolution, 
it should be easier to resolve emission regions of
pulsars located in high density environments such
as supernova remnants\cite{main+17b} or binaries in which the pulsar's
companion has an ionized outflow.
Here, we report events of extreme plasma lensing in the ``Black
Widow'' pulsar, PSR~B1957+20, near the phase in its 9.2 hour orbit in
which its emission is eclipsed by its companion's
outflow\cite{fruchter+88,fruchter+90,ryba+91}.  During the lensing
events, the flux is enhanced by factors of up to 70--80 at specific
frequencies.  The strongest events clearly resolve the emission
regions: they affect the narrow main pulse and parts of the wider
interpulse differently.
We show that the events arise naturally from density fluctuations in
the outer regions of the outflow, and infer a resolution of our lenses
comparable to the pulsar's radius, about 10\,km.
%
%Our results thus provide a physical scale of the emission regions, and offer the prospect of mapping their geometry.
%
Furthermore, the distinct frequency structures imparted by the lensing
are reminiscent of what is observed for the repeating fast radio burst
FRB 121102, providing observational support for the 
idea that this source is observed through, and thus at times
strongly magnified by, plasma lenses\cite{cordes+17}.

%suggesting the possibility that bursts from that source are also observed through, and thus at times strongly magnified by, plasma lenses.

\end{abstract}

% OBSERVATIONS / EMPIRICAL

\begin{figure*}
    \includegraphics[trim=0cm 0cm 0cm 0cm, clip=true,
    width=1.0\textwidth]{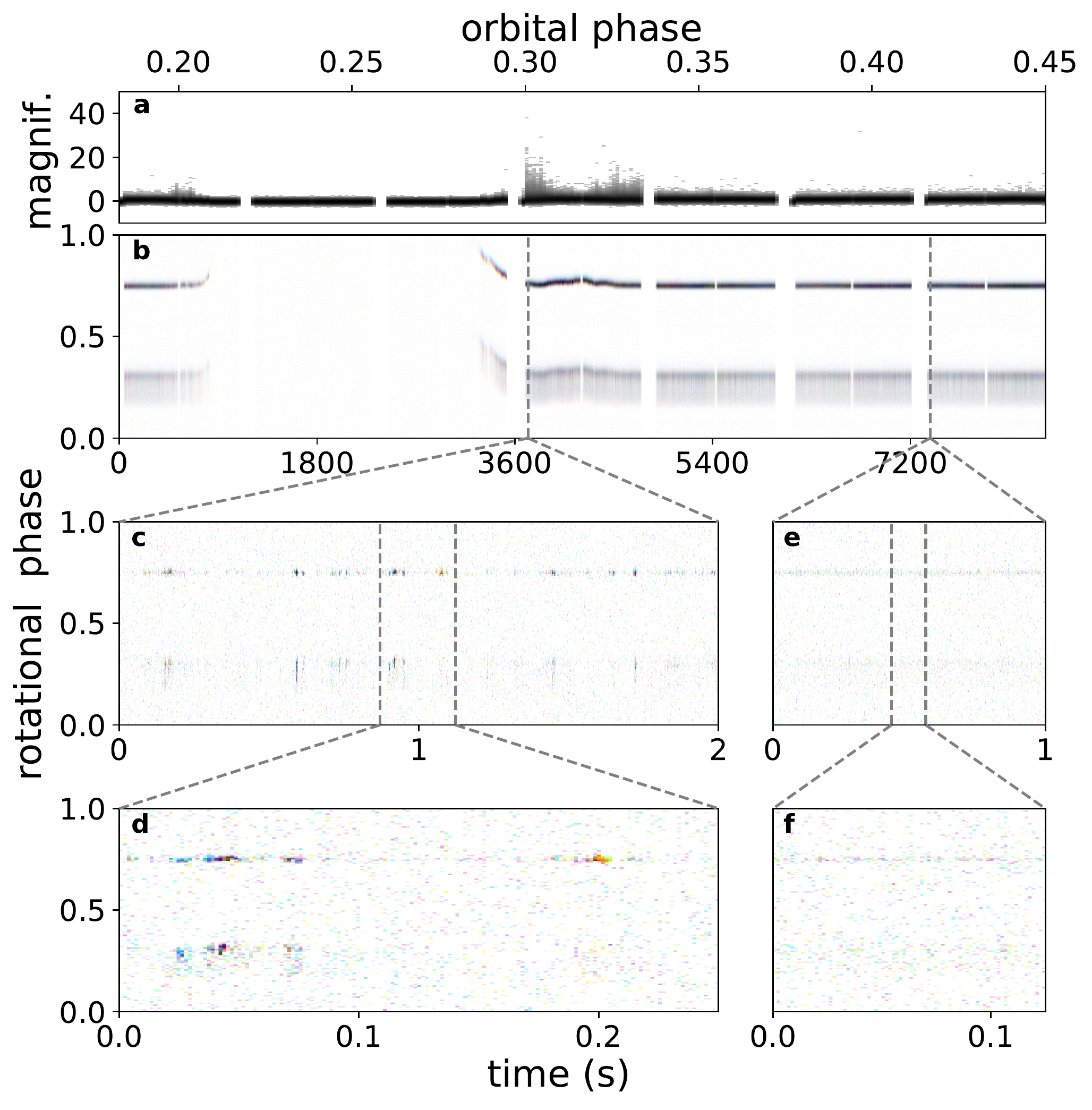}
    \caption{\textbf{Strong lensing of the pulsar.}
      \textbf{a}, Magnification distributions for the main pulse,
      in 10s time bins, showing periods of lensing near ingress and
      egress of the eclipse of the pulsar by its companion's outflow.
      The colorbar is logarithmic, to help individual bright pulses
      stand out.
      \textbf{b}, Pulse profiles as a function of
      time, averaged over $10\,$s and in 128 phase bins.  Cyan, yellow,
      and magenta represent contiguous
      $16\,$MHz sub-bands, from low to high frequency. Near eclipse,
      plasma dispersion causes frequency-dependent
      delays.  Gaps in the data correspond to calibrator scans.
      \textbf{c, d}, Enlargements in which each time
      bin corresponds to an individual, 1.6\,ms pulse. One sees extreme,
      chromatic lensing in which pulses are magnified by an order of
      magnitude over tens of ms; the brightest event is magnified by a
      factor of $\sim\!40$ across our highest frequency sub-band. In
      some events, the main pulse and interpulse are clearly affected
      differently, indicating that different emission regions
      corresponding to these are resolved by the lensing structures.
      \textbf{e, f}, Enlargements for a quiescent period for comparison.}
    \label{figure:LensingPanel}
\end{figure*}

On 2014 June 13-16, we took 9.5 hours of data of PSR B1957+20 
with the 305-m William E. Gordon Telescope at the Arecibo observatory, 
at observing frequency of 311.25--359.25\,MHz (see Methods). 
These data were previously searched for giant pulses\cite{main+17} -- 
sporadically occurring, extremely bright pulses, which are much shorter 
than regular pulses ($\lesssim\!1\,\mu$s compared to 10s of $\mu$s).
While we found many giant pulses at all orbital phases, we noticed that
the incidence rate of bright pulses was much higher leading up to
and following the radio eclipse.

As can be seen in Figure~1, most of the pulses
near eclipse do not look like giant pulses but rather like brighter
regular pulses -- they are bright over a large fraction of the pulse
profile, and most tellingly, occur in groups spanning several 1.6\,ms pulse
rotations, suggesting the underlying events last of order 10\,ms. 
Their properties seem similar to the hitherto
mysterious bright pulses associated with the eclipse of PSR
J1748$-$2446A\cite{bilous+11}, suggesting a shared physical mechanism.

High-magnification events are often chromatic, as can be seen from the 
colors in Figure~1 (which reflect 16\,MHz
sub-bands), and is borne out more clearly by the spectra shown in
Figure~2: some show frequency widths comparable to
our 48\,MHz band, peaking at
low or high frequency, while others show strong frequency evolution,
in some cases tracing out a slope in frequency-time space,
in others a double-peaked profile.

For many high-magnification events, the magnification is not uniform
across the pulse profile. The bottom panel of Figure~1 
shows this dramatically: at 0.2\,s, the
main pulse is greatly magnified over 5 pulses, while the entire
interpulse is barely affected.  In addition, the components of the
broad interpulse are often magnified differently from each other, as
can be seen most readily in the magnified events in Figure~3. 
Thus, the events resolve the pulsar's various emission regions.

\begin{figure}
  \centering
  \includegraphics[trim=1.5cm 3.3cm 6cm 1.5cm, clip=true, width=0.5\textwidth]{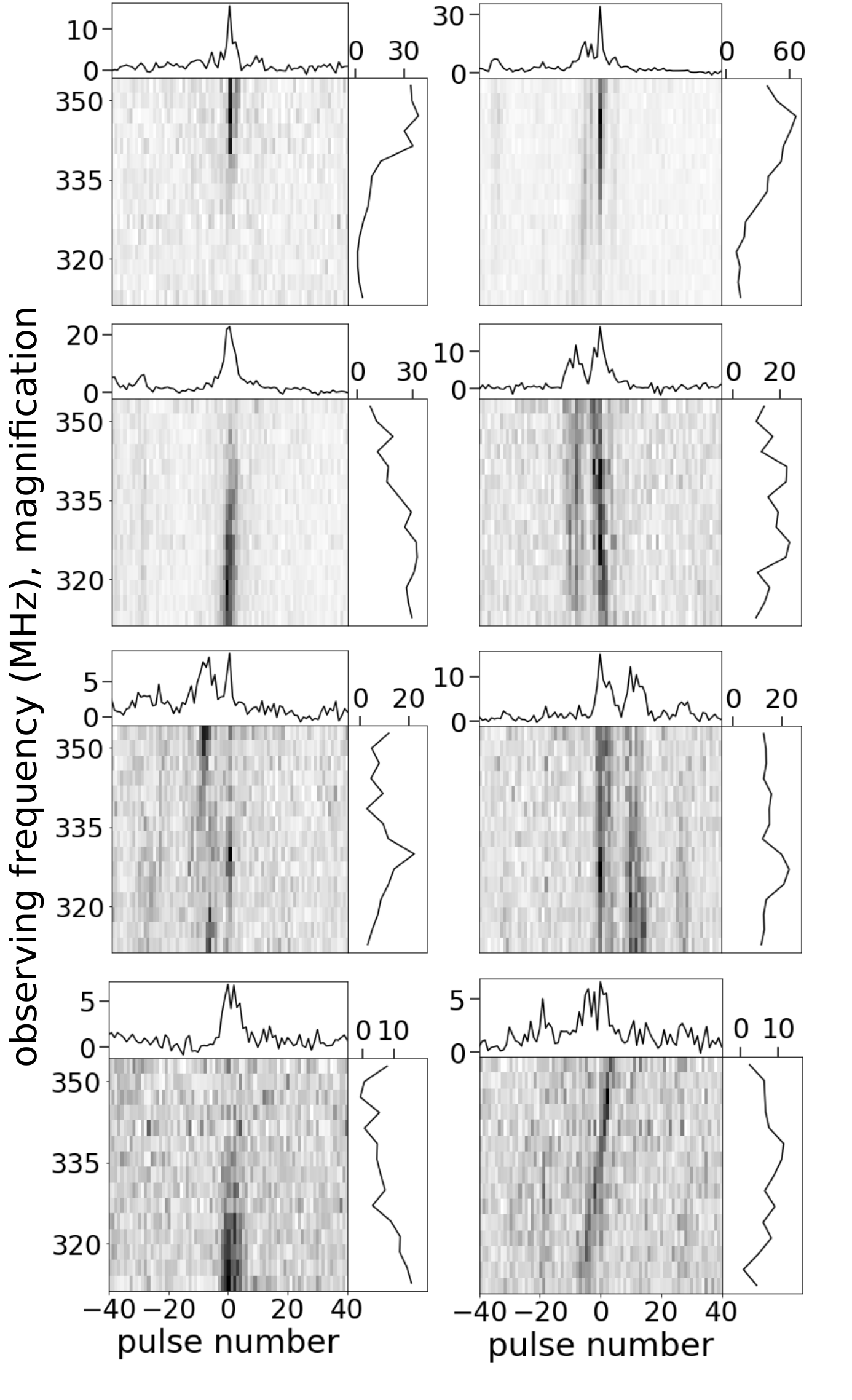}

  \caption{\textbf{A range of spectral behavior in lensing events.}  %
    Power spectra, binned to 3\,MHz, of the main pulse in
    consecutive 1.6\,ms rotations, for selected lensing events from eclipse egress.
    The top panels show the average magnification in each main pulse, and the side
    panels the magnification spectra of the brightest pulse (i.e., that of pulse
    number 0).}
  \label{figure:Spectra}
\end{figure}

\begin{figure}
  \centering
  \includegraphics[trim=0cm 0cm 0cm 0cm, clip=true,
  width=0.4\textwidth]{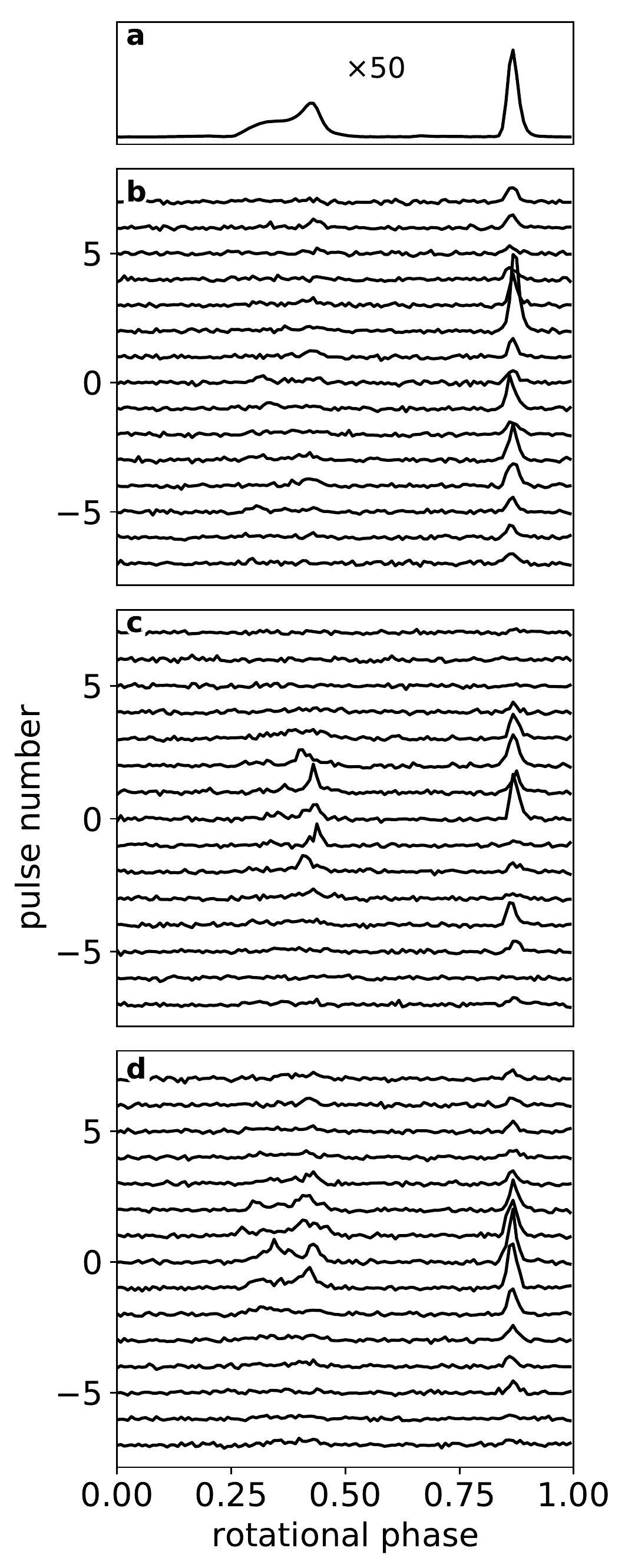}
  \caption{\textbf{Profiles of lensing events}.  %
    \textbf{b, c, d}, Frequency-averaged pulse
    profiles surrounding bright events (using 128 phase bins).  One
    sees that the events resolve the magnetosphere: the main pulse and
    interpulse are affected differently, as are parts within the
    interpulse.  \textbf{a}, Average pulse profile of a 9 minute quiescent
    region, scaled up by a factor of 50 for comparison purposes.}
\label{figure:Profiles}
\end{figure}

The most strongly magnified events occur within three specific time spans, 
each lasting for $\sim\!5$ minutes, around orbital phase $\phi = 0.20$,
$\phi = 0.30$, and $\phi = 0.32$ (see
Figure~1); here, $\phi = 0.25$ corresponds to
superior conjunction of the pulsar in its 9.2-hour orbit \cite{fruchter+88}
and the duration of the eclipses at 350\,MHz is about
40 to 60 minutes\cite{fruchter+88,fruchter+90,ryba+91}, or $\Delta\phi\simeq0.1$ 
(a sketch of the system geometry is shown in Extended Data Figure~1).
In each time span, we observe only small
overall frequency-dependent time delays, of order \SI{10}{\micro\second}, which
indicate modest increases in electron column density, with dispersion
measures (DM) of $\sim\!10^{-4}{\rm\,pc/cm^{3}}$ (see
Figure~1 as well as Extended Data Figure~3).  
Intriguingly, at times when the delays are longer
and the DM thus higher -- right before and after eclipse, and in between the
two post-eclipse periods of strong lensing, at $\phi\simeq0.31$
-- the magnifications are less dramatic, only up to a factor of a few,
and correlated over much longer timescales, of order 100\,ms.  After
the main lensing periods, up to $\phi\simeq0.36$, flux variations
remain correlated, suggesting even weaker events are still present
(examples of lensing in the aforementioned regions are shown
in Extended Data Figure~4).

% THEORY / INTERPRETATION

To determine whether the magnification events could be due to
lensing by inhomogeneities in the companion's outflow, we first
measured the excess delay due to dispersion at a time resolution of 2\,s,
the shortest at which we can measure it reliably (see
Methods).  We find that during the periods of strong magnification
events, the delay fluctuates by $\sim\!1\,\mu$s on this timescale.  Assuming
the relative velocity between the pulsar and the companion's outflow is
roughly the orbital velocity, $360{\rm\,km/s}$ 
(ref.~\cite{vankerkwijk+11}), the 2\,s timescale corresponds to a
spatial scale $\Delta x\simeq720\,$km.  For this scale, the expected
geometric delay is $\Delta x^2/2ac\simeq0.5\,\mu$s (where
$a=6.4\,$lt-sec is the orbital separation and $c$ the speed of light).  Since this is
comparable to the observed dispersive delays, lensing is
expected.

To model the magnified pulses, we treat the signal using a standard
wave optics formalism.  The electric field received by an observer
is the sum of the electric field of the source across the
lens plane, with the phase at every point determined by
both geometric and dispersive delays.  When a large area on the plane
has (nearly) stationary phase, the electric field combines coherently
leading to a strongly magnified image, with the magnification $\mu$
proportional to the area squared.  Since dispersive and geometric
delays scale differently with frequency $\nu$, a plasma lens cannot be in
focus over all frequency, but will impart a characteristic frequency
width.  This width will be smaller for larger magnification, with the
precise scaling depending on the extent to which the lens is elliptic:
$\Delta \nu / \nu \sim 1 / \mu$ for a very elongated, effectively
linear lens, and $\Delta \nu / \nu \sim 1 / \sqrt{\mu}$ for a
(roughly) circular one.  In strong lensing, one generically expects
multiple images to contribute, and thus caustics to form, which can lead to a
slope in time-frequency space, double-peaked spectra, as well as other
interference effects (see Methods).
All of these are consistent with the behavior seen in the measured
spectra of the magnified pulses in Figure~2.

% RESOLUTION - TWO ARGUMENT

For any lens, different pulse components will be magnified
differently if they arise from regions which have projected
separations larger than the lens's resolution.  We can make a
quantitative estimate of the lens resolution using the magnifications
of the events.  Continuing to assume the lens is roughly co-located
with the companion, i.e., at the orbital separation~$a$, the
resolution of the lens for given magnification would be
$\sim\!1.9R_1/\mu^{1/2}$ for a linear lens, or
$\sim\!1.9R_1/\mu^{1/4}$ for a circular one, where
$R_1\equiv \sqrt{\lambda a/\pi}$, or 23\,km at our observing
wavelength $\lambda = c/\nu \simeq90\,$cm (see Methods for a
derivation). A peak magnification of, eg., $\mu=50$, then corresponds
to a physical resolution of $\sim\!6{\rm\,km}$ for a linear lens, or
$\sim\!17{\rm\,km}$ for a circular one.

The inferred resolution is comparable to the $\sim\!10{\rm\,km}$ radius
of the neutron star, and substantially smaller than the light-cylinder
radius, $R_{\rm LC}\equiv cP/2\pi=76\,$km, where the velocity of
co-rotating magnetic fields approaches the speed of light, bounding
the magnetosphere in which pulsar emission is thought to originate\cite{ginzburg+75}.
Thus, the lensing offers the opportunity to map the emission geometry.

Qualitatively, since the main pulse and interpulse, as well as parts of the wide
interpulse beam, are sometimes magnified very differently, the inferred
resolution is a lower limit to both the projected separation between the
main pulse and the interpulse, and to the size of the interpulse.
Perhaps not unexpectedly, the spatial separations do not seem
to map directly to rotational phase: we see similar differences
between the main pulse and the interpulse, which are separated by half a rotation, and
between parts of the interpulse, which are separated by only $\sim$0.1 rotation.
This may indicate that the interpulse consists of multiple components, which are
not located close together in space.

Combining the lens resolution with the timescale of the lensing events
allows us to constrain the projected relative velocity between the
lens and the pulsar.  A priori, one might expect the outflow velocity
to be slow, in which case the relative velocity would just be the
orbital motion of $\sim\!360 \rm{km}\,\rm{s}^{-1}$.  Given that
strong magnification events typically last about 10\,ms, this would
then imply a resolution of $\sim\!4$\,km, not dissimilar from what we
find above, thus suggesting that the assumption of a slow outflow
velocity is reasonable.  By combining different constraints between
the duration and frequency widths of lensing events, we can set a quantitative
limit to the relative velocity (see Methods for a derivation), of
\begin{equation}
  v > 0.5\,\frac{R_1}{\Delta t_\text{HWHM}}
      \left(\frac{\Delta \nu}{\nu}\right)_\text{HWHM}^{1/2}
   \gtrsim 360\;{\rm\,km/s},
\end{equation}
where for the numerical value we use that the tightest constraints come from the
shortest, least chromatic events, which have frequency widths
$(\Delta \nu / \nu)_{\rm HWHM} \simeq 0.1$, and durations
$\Delta t_{\rm HWHM} \simeq 10{\rm\,ms}$ (measured as Half Width at
Half Maximum [HWHM]; see Figure~2).  This
approximate limit equals the relative orbital velocity, which implies
that the outflow velocity can be (but does not have to be) small in
the rest frame of the companion.

Our results offer many avenues of further research.  Ideally, one
would use the lensing to map the pulsar magnetosphere.
This requires better constraints on the lenses, e.g., from
observations over several eclipses at a range of frequencies.
Further observations may also shed light on the eclipse mechanism.  E.g.,
if it is cyclo-synchrotron absorption, large magnetic
fields, of $\sim\!20$\,G, are required\cite{thompson+94}, which would impart 
measurable polarization dependence of the lensing events.
Furthermore, combining density and velocity into a mass-loss
rate of the outflow, one can infer the life-time, and thus the final
fate, of these systems.

Finally, our observations establish that radio pulses can be strongly amplified
by lensing in local ionized material. This adds support for the proposal
that fast radio bursts (FRBs) can be lensed by host galaxy plasma,
leading to variable magnification, narrow frequency structures,
and clustered arrival times of highly amplified (and thus observable)
events\cite{cordes+17}. Evidence for high-density environments includes
that FRB 110523 is scattered within its host galaxy\cite{masui+15}, 
and that FRB 121102 has been localized to a star-forming
region\cite{tendulkar+17, bassa+17}, in an extreme and 
dynamic magneto-ionic environment\cite{michilli+18}.
Indeed, the latter's repeating bursts have spectra\cite{spitler+16,scholz+16,law+17}
remarkably similar to those shown in Figure~2
(e.g., compare with Figure~2 of ref.~\cite{spitler+16}), and, like those 
bursts, the brightest pulses in PSR B1957+20 are highly clustered in time. 

%\bibliography{b1957eclipse.bib}

%\printbibliography
%\printbibliography[check=onlynew]
%\printbibliography[segment=\therefsegment,check=onlynew,heading=subbibliography]

\begin{addendum}
    \item[Author Contributions]
		R.~M. discovered the magnified pulses; wrote the majority of the manuscript; and created the figures.
        I-S.~Y. developed and wrote the sections on wave optics formalism in Methods;
        M.~H.~v.~K and U-L.~P. guided the analysis and interpretation of the results;
        M.~H.~v.~K. also helped write the manuscript and influenced the content and presentation of the figures;
        F-X.~L., V.~C., and K.~V. quantified the excess DM associated with the eclipse. 
        N.~M. wrote the code used to produce the coherently de-dispersed single pulse profiles used in this analysis, and derived an improved orbital ephemeris relative to which time delays are measured.
        D.~L. worked on criteria to systematically separate giant pulses from pulses magnified by plasma lensing.
        All authors contributed to the interpretation of results.
        
    \item[Competing Interests] The authors declare that they have no competing financial interests.
    \item[Correspondence] Correspondence and requests for
        materials should be addressed to R.~M. (email: main@astro.utoronto.ca).
\end{addendum}

\textbf{Acknowledgements.}
We thank A. Bilous for sharing and discussing her results which motivated
this analysis. We thank the rest of our scintillometry group for discussions
and support throughout. R.M., M.H.v.K., U.-L.P. and K.V. are funded
through NSERC. I-S.Y was supported by a SOSCIP Consortium
Postdoctoral Fellowship.

\clearpage

\begin{methods}

\subsection{Calculating magnifications.}
The data were taken in four 2.4\,hr sessions on June 13--16, 2014, and
were recorded as part of a European VLBI network program (GP~052). The
data span 311.25--359.25\,MHz, in three contiguous 16\,MHz sub-bands,
and we read, dedispersed (with a dispersion measure of $29.1162{\rm\,pc/cm^3}$),
and reduced the data as described in
ref~\cite{main+17}.  As one extra step, we accounted for the
wander in orbital period of PSR B1957+20 by adjusting the time of
ascending node in the ephemeris, such that it minimized the scatter of
the arrival times of giant pulses associated with the main pulse
across all four days of observation.

We define the magnification of a pulse as the ratio of its flux to the
mean flux in a quiescent region far from eclipse. Specifically, we
construct an intensity profile of each pulse using 128 phase gates,
subtract the mean off-pulse flux in each 16\,MHz sub-band separately,
measure the flux in an 8-gate ($\sim\!100{\rm\,\mu s}$) window around
the peak location (which we find from folded profiles, averaged in
2\,s bins to obtain sufficient signal-to-noise), and divide by the
average flux in the same pulse window measured in a 9-minute section
far from eclipse.

To construct the spectra of lensing events, we start by binning power
spectra in 3\,MHz bins.  We correct these approximately for the
bandpass and the effects of interstellar scintillation (which still
has a small amount of power on 3\,MHz scales) by dividing them by the
average spectra in the 15 seconds before and after (excluding the
lensing event itself).  This time span is chosen to be safely less
than the timescale of $\sim\!84\,$s on which the interstellar
scintillation pattern varies \cite{main+17}, ensuring the dynamic
spectrum is stable on these scales. With 30\,s of data, it is also
very well measured: each 3\,MHz channel has $S/N \simeq 150$.

\subsection{Evidence that strong plasma lensing must occur.}
To obtain the properties of the lensed images, and in particular to
determine whether we are in the ``strong'' or ``weak'' lensing regime
-- i.e., whether or not multiple images are formed -- we use the basic
principles of wave optics, considering path integrals of the electric
field over a thin lens \cite{born+99}.  In our case, the phase of an
electromagnetic wave going through different paths has contributions
from both geometric and dispersive time delays,

\begin{eqnarray}
  \phi(x,y) = \phi_{\textrm{GM}}(x,y) + \phi_{\textrm{DM}}(x,y),
\end{eqnarray}
where $x$ and $y$ are in the lens plane.  For a geometrically thin
lens, i.e., with thickness along the line of sight much smaller than
the separation $a$, the geometric contribution to the phase can be
written as,
\begin{equation}
  \phi_{\textrm{GM}} = \frac{1}{2}\left(\frac{x^2+y^2}{a^2}\right)
                       \left(\frac{a}{\lambda}\right)
                     = \frac{x^2+y^2}{2 R_{\rm{Fr}}^{2}},
  \label{eq-phiGM}
\end{equation}
where $R_{\rm Fr}$ is the Fresnel scale,
\begin{equation}
  R_{\rm{Fr}} \equiv \sqrt{\lambda a} \simeq 40{\rm\;km},
\end{equation}
with $\lambda$ the wavelength of the radiation.  For the numerical
value, we used $\lambda = c/\nu = 90{\rm\,cm}$ (where $\nu$ is the
observing frequency) and assumed that the lensing material was
associated with the companion and thus at the orbital separation of
$a\simeq6.4\,$lt-sec. (Note that by using wavelengths and frequencies
of full cycles, our phases are in cycles too, not radians.)

The dispersive contribution arises from the signal propagating through
the lens's extra dispersion measure (electron column density)
$\Delta DM=\int n_e\,dz$ (with $n_e$ the electron number density),
\begin{equation}
  \phi_{\textrm{DM}}(x,y) = -\frac{k_\text{DM}}{\nu} \Delta DM(x,y),
\end{equation}
where $k_{\textrm{DM}} = e^2/2\pi m_e c =
4148.808\rm{\,s\,pc^{-1}\,cm^3\,MHz^{2}}$ (ref.~\cite{manchester+72}).
The minus sign arises because in a
plasma the phase velocity is greater than the speed of light.

Integrating over different paths, one effectively select regions where
the electric fields are coherent; these lead to the final images.  For
instance, if the extra electrons are distributed uniformly (in the
$x$\nobreakdash--$y$~plane after the $z$ integral), then
$\phi_{\textrm{DM}}$ is approximately constant and the total phase has
a stationary point around $x=y=0$, i.e., around the line of sight.
Furthermore, all paths which are less than the Fresnel scale away from
this central path have similar phase, and thus one recovers the
general result that the area of the lens that contributes scales with
$R_{\rm Fr}^2$.

For the lensing to be strong, a minimum requirement is that changes in
the geometric and dispersive phase are of similar magnitude.  Since
dispersion also leads to overall time delays, differences in pulse arrival
time give a handle on inhomogeneities in the lensing material.
We measure pulse arrival times using the usual
procedure of fitting pulse profiles to a high signal-to-noise template
(from a quiescent period).  We fit profiles in the three bands separately,
and convert the weighted average to $\Delta DM$ using standard equations.
We find that the inferred $\Delta DM$ shows both the expected
large-scale variations\cite{fruchter+88} and significant 
variability down to the shortest time scales, of 2\,s, at which we can
measure it reliably (see Extended Data, Fig.3).  During the
periods in which the strongly magnified events occur, we find intrinsic
variability corresponding to variations in delay of
$\Delta t_{\rm DM} \simeq 1{\rm\,\mu s}$, which suggests the
inhomogeneities correspond to differences in the dispersion phase of
$\Delta\phi_{\rm DM} = \nu \Delta t_{\rm DM} \sim 300{\rm\,cycles}$
at 2\,s timescales.

To compare this to the geometric phase, we need to translate the
timescale of 2\,s to a length scale.  Assuming the relative motion
between the pulsar and the lensing material is dominated by the
orbital motion, of $360{\rm\,km/s}$, the spatial scale is
$720{\rm\,km}$.  Using Eq.~(\ref{eq-phiGM}), this corresponds to
$\Delta\phi_{\textrm{GM}}(720{\rm\,km}) \sim 200{\rm\,cycles}$.

The fact that $\Delta \phi_{\textrm{DM}}$ is comparable and slightly
larger than $\Delta\phi_{\textrm{GM}}$ at the $720\,\rm{km}$ distance
scale guarantees that somewhere around this scale, slopes in
$\phi_{\textrm{DM}}$ and $\phi_{\textrm{GM}}$ will sometimes cancel.
It also implies that multiple stationary points with $\nabla\phi=0$
are likely.  Since the phases result from a continuous function of $x$,
$y$ and $\nu$, the stationary points must emerge or annihilate in
pairs, leading to so-called caustics, where one has not just
$\nabla\phi=0$ but also
${\rm det}(\partial_i\partial_j\phi) = 0$\:\cite{nye99}.

Given the above, strong lensing is a natural cause of the strongly
magnified events we observe.

\subsection{A perfect lens model.}
To calculate properties of the lensed images, we would need to perform
the path integral.  This is not possible since it requires the value
of $\phi_{\textrm{DM}}$ of the full two-dimensional lens plane at a
resolution better than the Fresnel scale, while our time-delay
measurements are limited to the one-dimensional trajectory of the
pulsar, and to an order of magnitude larger scales.

Hence, for our analysis we use a simplified model, amenable to 
wave-optics analysis, in which each strong lensing event
is associated with an elliptical perfect lens, i.e., there is a single
focal point to which all paths within the lens contribute perfectly
coherently, and all paths outside the lens do not contribute at all.
As shown in Extended Data Figure~2, our model
is parametrized by the location of the focal point relative to the
centre of the lens, $(X,Y)$, the semi-major and semi-minor axes of the
lens, $(\Delta x_{\rm lens}, \Delta y_{\rm lens})$, and the
orientation relative to the projected pulsar trajectory.

In the special case of a centred circular lens with radius $R$, i.e.,
$\Delta x_{\rm lens}=\Delta y_{\rm lens}=R$ and $X=Y=0$, the setup
would produce an Airy disk.  By comparing this to the actual
path-integral of an unlensed image (i.e., with
$\phi_{\textrm{DM}}=0$), we can identify the unlensed case with a
circular lens with radius $R_1\equiv R_{\rm Fr}/\sqrt{\pi}$.

Integrating the electric field over an elliptical lens, the
magnification (defined as the increase in intensity
$\mu = I / \langle I \rangle$, where $I = E^{2}$), will be
proportional to the area squared,
\begin{equation}
\mu = \left(\frac{\Delta x_{\rm lens}}{R_1}\right)^2
      \left(\frac{\Delta y_{\rm lens}}{R_1}\right)^2.
      \label{eq-mu}
\end{equation}

Before continuing, we note that at first glance it might appear
puzzling for the magnification to be proportional to image area
squared, since that seems to violate energy (flux) conservation.
However, the image is also more highly beamed; what we calculated is
the magnification in the center of the beam.  Viewed from an angle, a
linear phase term is induced across the originally coherent region.
Thus, when the region is larger, it is easier to become incoherent.
The solid angle scales as
$\Delta\Omega \propto \Delta x_{\rm lens}^{-1} \Delta y_{\rm
  lens}^{-1}$, and thus total energy (flux) is indeed proportional to
the area of the lens.

\subsection{Differential magnification and chromaticity.}
We now proceed to derive the physical resolution of the lens.  The
above magnification is reached only when the source is exactly at the
focal point.  When the emission region is some distance away from the
focal point, paths from it to different parts of elliptical region
will have extra phase differences.  For instance, when a source is
separated from the focal point in the $x$ direction by
$x_{\rm source} \equiv (X-x_s)$ (where $x_s$ is measured relative to
the centre of the lens), there is a phase difference across the lens
of
\begin{equation}
\Delta\phi = \frac{\Delta x_{\rm lens} ~ x_{\rm source}}{R_1^{2}}.
\label{eq-shift}
\end{equation}
When this phase difference reaches order one, the image from the new
source location will no longer be magnified.  Defining the resolution
as the offset for which total cancellation happens, we find from an
explicit integral over the elliptical lens,\footnote{With a simple
  rescaling, the elliptical integral can be written as the circular
  one leading to an Airy disk.}
\begin{eqnarray}
  x_{\rm res} &\simeq& \frac{1.9R_1^2}{\Delta x_{\rm lens}},\\
  y_{\rm res} &\simeq& \frac{1.9R_1^2}{\Delta y_{\rm lens}}.
\end{eqnarray}
Hence, total cancellation occurs on an elliptical beam with semi-minor
and semi-major axes $x_{\rm res}$ and $y_{\rm res}$ that scale inversely
to the semi-major and semi-minor axes of the lens.

We can write the above in terms of the magnification for a specific
lens model.  We will consider two extremes, the first a very
anisotropic lens, effectively one-dimensional, for which $\Delta
y_{\rm lens}=R_1$.  For this case,
\begin{equation}
x_{\rm res} \simeq \frac{1.9 R_1}{\mu^{1/2}}.
\end{equation}
In the opposite direction, we consider a circular lens, with $\Delta
x_{\rm lens} = \Delta y_{\rm lens}$.  For this case,
\begin{equation}
x_{\rm res} = y_{\rm res} \simeq \frac{1.9 R_1}{\mu^{1/4}}.
\end{equation}
Here, dividing by the distance $a$ and writing in terms of diameter
$D=2\Delta x_{\rm lens}$, one recovers the usual relation for the
angular resolution of a circular lens, $1.22\lambda/D$.

A similar result can be derived in frequency space.  Given the
different frequency dependencies of the geometric and dispersive phases,
a lens cannot be fully in focus across all frequencies, but will have
some characteristic frequency width.  To show this, we first assume
that the focal point is co-located with the centre of the elliptical
lens, i.e., $(X,Y)=(0,0)$, and that perfect coherence occurs at some
frequency $\nu_c$, i.e., that at $\nu_c$ one has,
\begin{equation}
\phi_{\textrm{GM}}(x,y,\nu_c) = - \phi_{\textrm{DM}}(x,y,\nu_c) = \frac{\nu_c (x^2+y^2)}{2ac}.
\end{equation}
The total phase at a nearby frequency is then given by
\begin{equation}
\phi(x, \nu_c+\Delta \nu) = \left(\frac{\nu_c+\Delta \nu}{\nu_c}
                          - \frac{\nu_c}{\nu_c+\Delta \nu}\right)
                          \frac{\nu_c (x^2+y^2)}{2ac},
\end{equation}
the dominant term of which is linear in $\Delta \nu$.  If this term is
of order one within the elliptical region, the image will no longer be
magnified.  Perfect cancellation does not happen in this case, but one
can define characteristic width.  Using half-width at half-maximum,
one finds,
\begin{equation}
  \left(\frac{\Delta \nu}{\nu_c}\right)_{\textrm{HWHM}} =
  \frac{2.9 R_1^2}
       {\sqrt{ 3\Delta x_{\rm lens}^4
              -2\Delta x_{\rm lens}^2 \Delta y_{\rm lens}^2
              +3\Delta y_{\rm lens}^4 }}.
  \label{eq-dfmu}
\end{equation}

In terms of the magnification, we find for the special case of an
effectively one-dimensional lens,
\begin{equation}
  \left(\frac{\Delta \nu}{\nu_c}\right)_{\textrm{HWHM}} \simeq \frac{1.7}{\mu},
\end{equation}
while for a circular lens,
\begin{equation}
  \left(\frac{\Delta \nu}{\nu_c}\right)_{\textrm{HWHM}} \simeq \frac{1.5}{\mu^{1/2}}.
\end{equation}

When the focal point is not in the center of the elliptical region,
there will be linear terms proportional to $X \Delta\nu$ and
$Y\Delta\nu$ that also contribute to the phase.  They make the
dependence on $\Delta \nu$ even steeper, which means that
Eq.~\ref{eq-dfmu} is an upper bound on the frequency width of
magnified images.

It is possible for the frequency and position shifts to cancel each
other, so that after moving the source by some distance, it is still
strongly magnified at a nearby frequency.  That leads to a slope of
strongest magnification in time-frequency space.  Following a 
derivation similar to that outlined above for the behavior with frequency and
position separately, we find that the slope is related only to the
offset of the source from the focal point. For instance, for the case 
that the source is traveling along the semi-major axis of the lens, i.e.,
along a trajectory $(x_{\rm s}, y_{\rm s})=(x_{\rm s}(t), 0)$, we find
\begin{equation}
\frac{d\nu}{dx_{\rm s}} = \frac{\nu_c}{2x_{\rm s}}.
\end{equation}

Finally, it is possible that the pulsar is close to the focus of more
than one lens at the same time.  As the pulsar moves, these multiple
focal points lead to multiple images images which change intensity and
interfere with each other.  This can lead to a wide variety of
spectral behavior, and might well be responsible for the multi-peaked
structures seen in some panels of Figure~2.

\subsection{A lower bound on the relative velocity.}
Within the perfect lens approximation, one can derive a lower bound on
the relative, transverse velocity between the lens and the pulsar using
the durations and frequency widths of the lensing events.

The duration of strongly magnified events is predicted to be,
\begin{equation}
  \Delta t_{\textrm{HWHM}} \simeq
      \frac{0.7 R_1^2}
           {\sqrt{\Delta x_{\rm lens}^2 v_x^2 +
               \Delta y_{\rm lens}^2 v_y^2}}.
  > \frac{0.7 R_1^2}{v \Delta x_{\rm lens}},
  \label{eq-dtobs}
\end{equation}
where the inequality corresponds to making the most conservative
estimate, that the lens is effectively one-dimensional, extended
in the $x$ direction, and that the full velocity $v$ is directed along it.

In the frequency space, the same event will have its width bounded
from above by Eq.~(\ref{eq-dfmu}),
\begin{equation}
\left(\frac{\Delta \nu}{\nu}\right)_{\textrm{HWHM}}
  < \frac{1.8 R_1^2}{\Delta x_{\rm lens}^2}.
\end{equation}

Combining the two limits to eliminate $\Delta x_{\rm lens}/R_1$, we find,
\begin{equation}
  v > 0.5\frac{R_1}{\Delta t_{\textrm{HWHM}}}
         \left(\frac{\Delta \nu}{\nu}\right)_{\textrm{HWHM}}^{1/2}.
\end{equation}

\subsection{Lensing consistency checks.}
Our analysis in terms of lensing was motivated by the fact that the
bright pulses before and after eclipse do not look like giant pulses, 
that they occur at specific orbital phases where DM fluctuations are large,
and that, in the majority of events, the
entire profile increases in flux (although not always by the same
factor across the profile), with the enhancements lasting for several
pulses.

Consistent with the expectations of plasma lensing, in which lenses
focus light but arrive with associated ``shadows'' where light is
scattered away from our direct line of sight, we find that in regions
with many strongly lensed pulses, the average flux received is
unchanged (see Extended Data Figure~4).
Also consistent with plasma lensing is that the strongly
magnified pulses are highly chromatic, often peaking at high or low
frequencies, and sometimes showing slopes in frequency time space or
interference patterns characteristic of caustics.

Finally, as a concrete example, the peak magnifications of
$\mu \sim 70$ are comparable to the magnifications expected for plasma
lensing.  The measured dispersive time delay changes provide direct
evidence that the geometric and dispersive phases $\phi_{\textrm{GM}}$
and $\phi_{\textrm{DM}}$ can cancel over distances of $\sim\!720$\,km.
The cancellation will likely happen mostly in one spatial direction;
setting $\Delta x_{\rm lens}=720$ km and $\Delta y_{\rm lens}=R_1$,
our perfect lens model predicts a magnification of order a hundred,
comparable to the observed value.

\textbf{Data Availability.}  The data underlying the figures is available in text files at 

$https://github.com/ramain/B1957LensingData$

\textbf{Code Availability.}  The raw data were read using the baseband package: 

$https://github.com/mhvk/baseband$

\end{methods}

\section*{Extended Data}

\setcounter{figure}{0}
\captionsetup[figure]{name=Extended Data Figure}

\begin{figure*}
\centering
\includegraphics[trim=3cm 5cm 6cm 6cm, clip=true, width=0.8\textwidth]{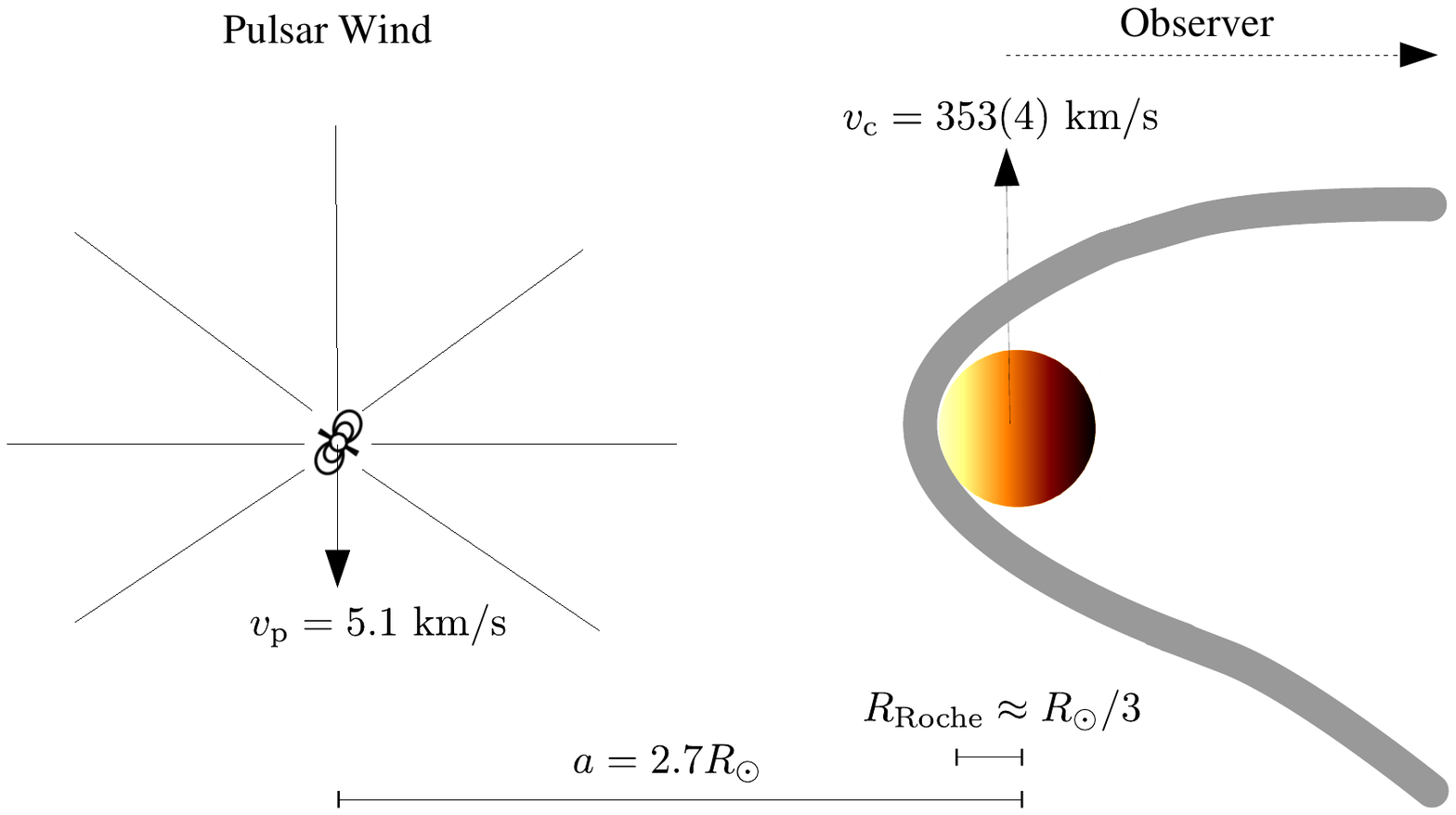}
  
  Extended Data Figure 1: \textbf{Shock geometry.} The brown dwarf companion is
  irradiated by the pulsar wind, causing it to be hotter on the side
  facing the pulsar \cite{vanParadijs+88}, and inflated, nearly
  filling its Roche lobe
  \protect{\cite{reynolds+07}}. Outflowing material is shocked by the
  pulsar wind, leaving a cometary-like tail of material. 
  This tail is asymmetric because of the companion's
  orbital motion, which leads to eclipse egress lasting substantially longer than
  ingress. The companion and separations are drawn roughly to scale,
  while the pulsar is not -- the light cylinder radius of 76 km would
  be indistinguishable on this figure.  The inclination of the system
  is conservatively constrained to $50 < i < 85\deg$ (ref.~
  \protect{\cite{vankerkwijk+11}}).
\label{figure:ShockDiagram}
\end{figure*}

\begin{figure*}
\centering
\includegraphics[trim=2cm 6cm 3cm 3cm, clip=true, width=0.8\textwidth]{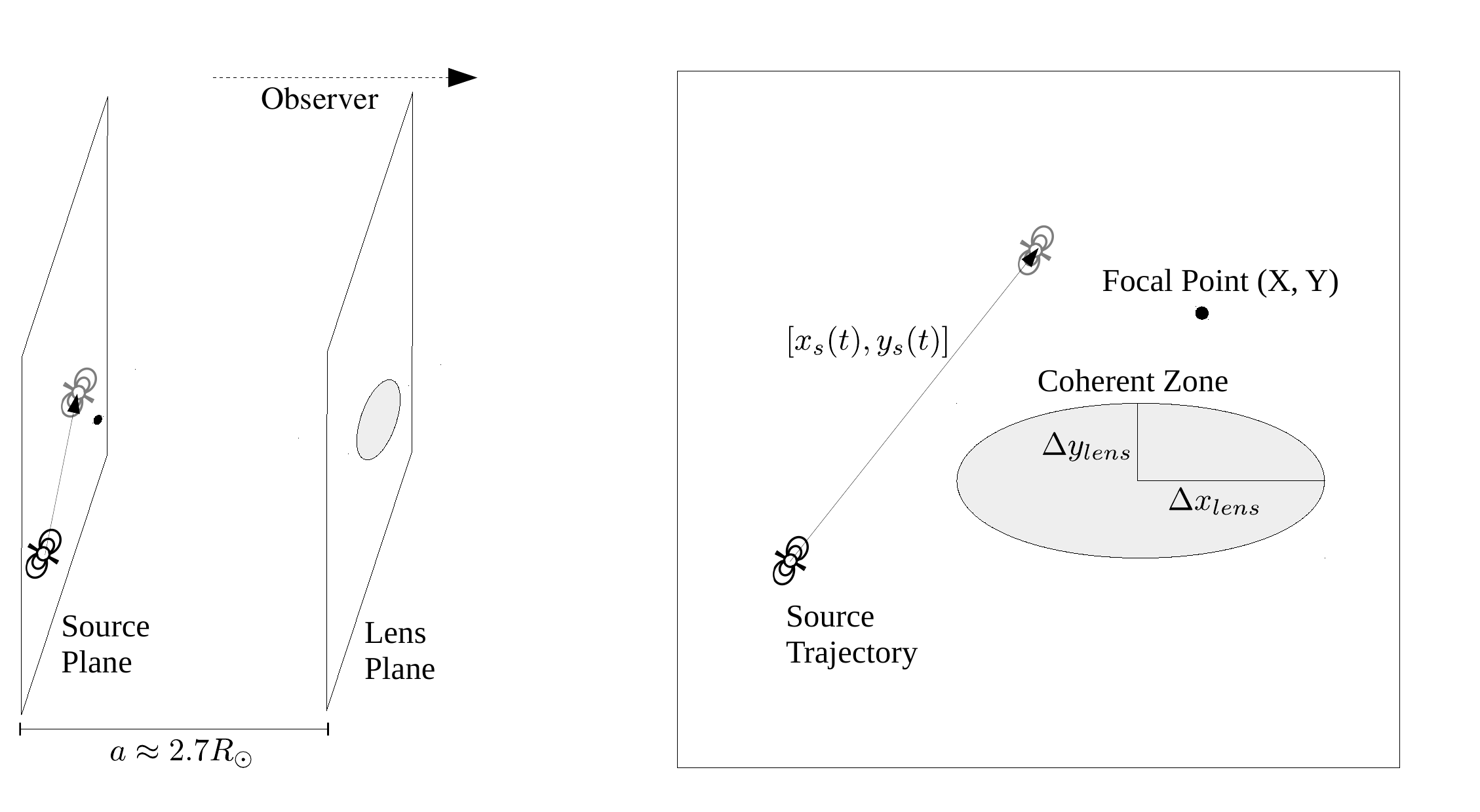}

Extended Data Figure 2: \textbf{Geometry of a lensing region}. An
  almost edge-on (left) and a face-on (right) view of the lensing
  geometry.  We assume that the source is at a separation equal to the
  semi-major axis $a$ of the binary system and moves on a
  trajectory parallel to the lens plane.  A source at the focal point
  $(X,Y)$ will illuminate the entire elliptical lens coherently,
  leading to strong magnification.  In general, the focal point may not
  be at the center of the
  ellipse (although it is likely within it), and the source
  trajectory $[x_s(t), y_s(t)]$ may not intersect the focal point or the
  elliptical region.
\label{figure:LensDiagram}
\end{figure*}

\clearpage

\begin{figure*}
\centering
\includegraphics[trim=0cm 0cm 0cm 0cm, clip=true, width=1.0\textwidth]{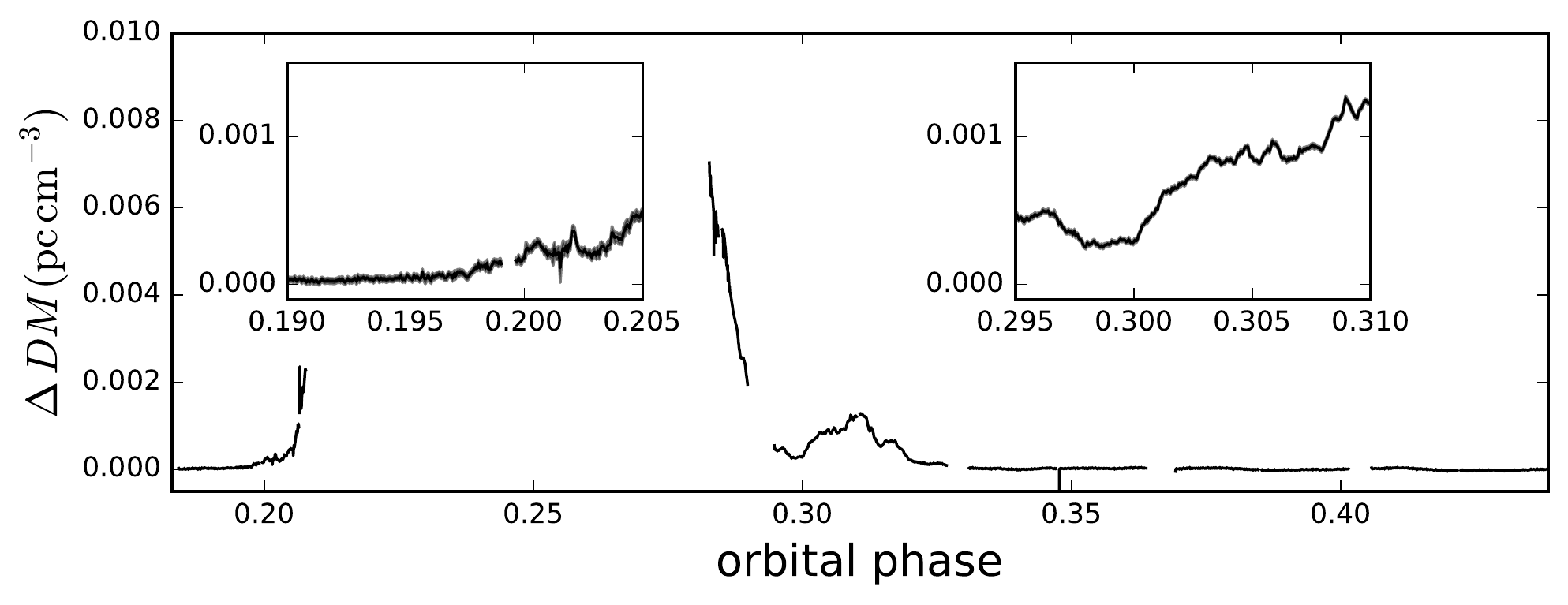}
Extended Data Figure 3: \textbf{Dispersion measure near the radio eclipse.} Shown is
  the excess dispersion measure relative to the interstellar
  dispersion, with insets focusing on regions of strong lensing.  The 
  excess dispersion is estimated from delays in pulse arrival times
  (see Methods); at our observing frequency of $330\,$MHz,
  $\Delta DM=0.001{\rm\,pc\,cm^{-3}}$ corresponds to a delay of
  $38\,\mu$s.  The scatter around the curves is intrinsic.  During
  periods of strong lensing, it is at a level
  of $2.6\times10^{-5}{\rm\,pc\,cm^{-3}}$, corresponding to variations in
  delay time of $1\,\mu$s.
\label{figure:DM}
\end{figure*}

\begin{figure*}
\centering
\includegraphics[trim=0cm 0cm 0cm 0cm, clip=true, width=0.8\textwidth]{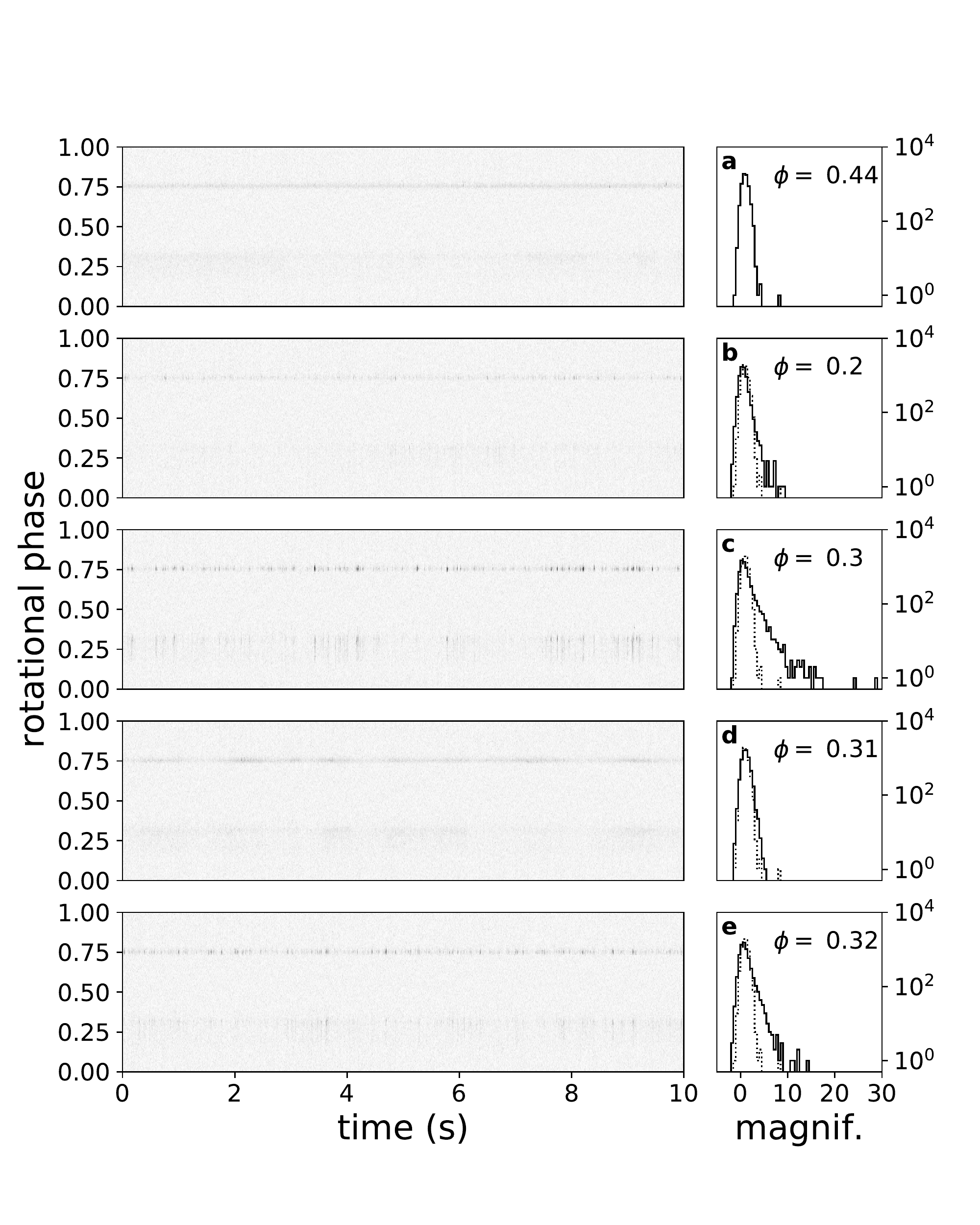}

Extended Data Figure 4: \textbf{Pulse profiles and magnification distributions.}
  Each row shows pulse profiles for 10\,s segments (left, in 128 phase bins),
  and the corresponding magnification distributions.  The segments are taken from:
  \textbf{a}, a quiescent region, showing a log-normal magnification
  distribution (repeated using a dotted line in \textbf{b}--\textbf{e} for comparison);
  \textbf{b}, eclipse ingress, showing some lensing events, reflected in the
  tail to high magnification; although hard to see, the average flux is reduced 
  to $\sim 70\%$ by absorption in eclipsing material);
  \textbf{c}, the first post-eclipse lensing period, showing extreme magnifications
  but little change in average flux;
  \textbf{d}, in between the two post-eclipse strong lensing periods, showing only
  weak lensing on relatively long, $\sim\!100\,$ms timescales;
  \textbf{e}, the second post-eclipse period of strong lensing.
\label{figure:LensingPanels}
\end{figure*}

\end{document}